\documentstyle[epsfig]{article}
\begin{document}


\section*{Quark Matter 99 Summary: Hadronic Signals}
\textbf{R. Stock\\
Physics Department, University of Frankfurt, D-60486 Frankfurt}

\vspace{0.5cm}

\section{Introduction}

It was obvious at this conference that the heavy beam program
at the AGS ($Au$) and at the SPS ($Pb$) has been completely successful,
all experiments delivering the physics that they were built for.
We have been presented
such a wealth of beautiful precision data (from the AGS in particular)
that no summary can give proper credit to everybody. Neither
is the restriction to ''hadronic signals'' helpful in reducing
the scope because, with the exception of direct photons and lepton
pair continua we have been discussing hadron production throughout.
In fact the long standing discussion of threshold effects in
charmonium production, upon variation of impact parameter and/or
system size has inspired a comprehensive program of similar searches
for ''onset'' phenomena  in bulk hadron production data. This
will be my first major topic: look at the evolutions with system
size (from $p+p$ via either intermediate mass nuclei or semiperipheral
collisions to central mass 200 collisions) and with $\sqrt{s}$. I will
then turn to the progress
in understanding the differences between the two major decoupling
stages during expansion: first from inelastic interactions (''chemical
freezeout''), lateron from resonance decay and elastic interaction
(''global freezeout''). Starting with the latter the progress
concerning the evidence for collective transverse velocity fields
will be reviewed. In the final section I will discuss how the
chemical freezeout stage (that fixes the hadronic production
rates and their ratios) is intimately related to the QCD hadronization
process, leading to the conclusion that analysis of such data
in central $Pb+Pb$ collisions at
$\sqrt{s}=17 \: GeV$ reveals the position
of the parton to hadron phase transformation in temperature,
energy density and baryochemical potential.

\section{New types of data}

Before engaging with the heavy systematics announced above I
wish to mention a few highlights of data pointing beyond the
presently well-discussed physics. Of course a strictly subjective
choice, with advance apologies to the unavoidable omissions.

Fig.~1 shows the combined WA98 [1] and NA45 [2] data for identified
pions in central $Pb+Pb$  collisions at top SPS energy. The transverse
mass scale ranges up to $4\: GeV$, far beyond the domain in $m_T$ on
which we normally base our arguments concerning thermal or radial
flow model inspired inverse slopes. Actually the local inverse slope
parameters vary throughout the scale due to the overall concaveness,
a warning with regard to flow model interpretation of pion spectral
data but, on the other hand, perhaps the first glance at hard partonic
scattering in SPS $A+A$ collisions - a topic emphasized by Gyulassy
[3] for future RHIC physics of jet attenuation etc.. We thus
expect to see these data in a new light at QM 2000.
\begin{figure}
\begin{center}
\epsfig{figure=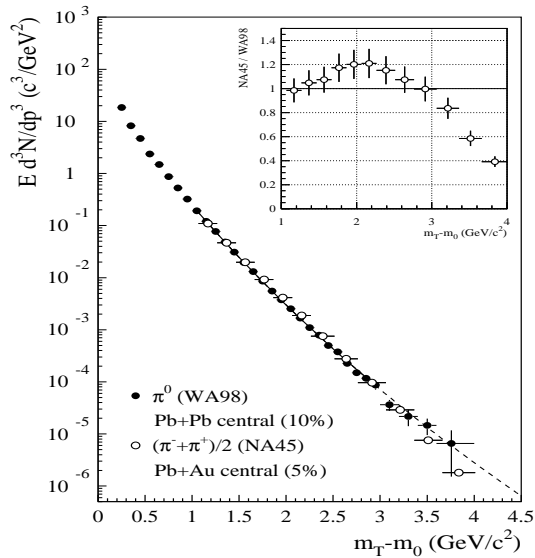,height=7.5cm, width=7.5cm}
\caption{Identified pion transverse mass spectrum from
WA98 [1] and NA45 [2] for central $Pb+Pb$ collisions at 158 $GeV/A$.}
\end{center}
\end{figure}

Turning to pion Bose-Einstein correlations the review talk by
Wiedemann [4] discussed the very indirect way in which straight
forward geometrical source size reflects in the data. But some
confidence in a naive geometrical interpretation is regained
from E895 results for semi-central $Au+Au$ collisions [5]. Fig.~2
shows the data for the ''sideward radius'' $R_s$ which was determined
(for the first time) differentially, in azimuthal orientation
relative to the eventwise reaction plane. Intuitively the reaction
fireball should look large as it is seen head-on, and small(er)
looking edge-on; and it does. Further evidence that HBT analysis
can be applied to single events and their source geometry (and
not only to ensemble averages of thermally coherent subvolumes,
terms with which the hydrodynamical models [6] frighten the practitioner)
is given by Fig.~3.
\begin{figure}
\begin{center}
\epsfig{figure=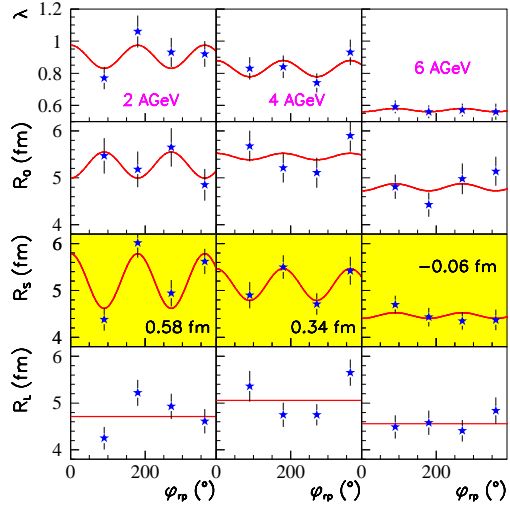,height=7cm, width=7cm}
\caption{HBT pion pair analysis by E895 [5] of semiperipheral
$Au+Au$ collisions, showing dependence of source parameters
relative to the eventwise reaction plane.}
\end{center}
\end{figure}

\begin{figure}
\begin{center}
\epsfig{figure=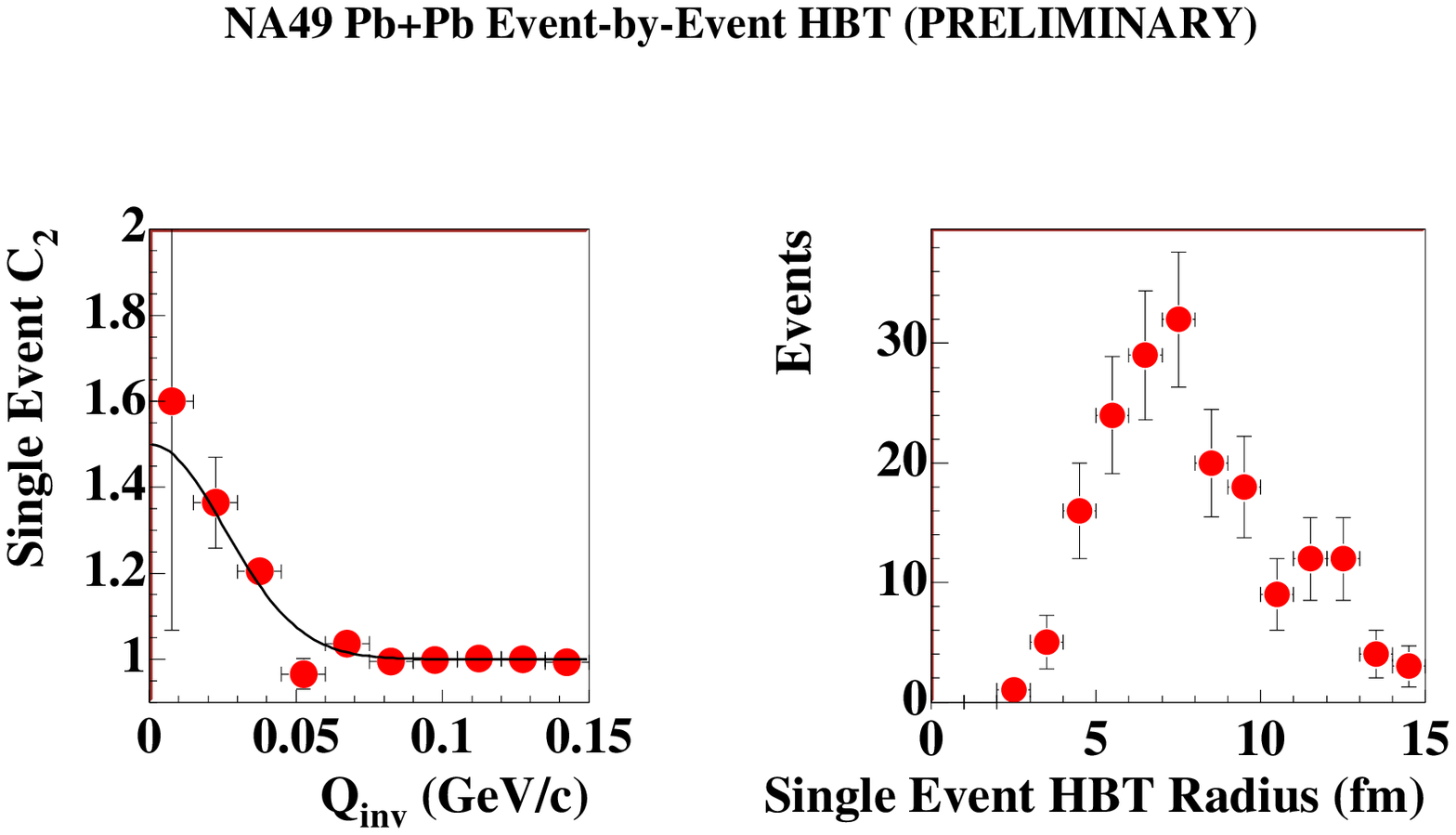,height=10cm, width=14cm}
\caption{Event by event HBT analysis of central $Pb+Pb$ by NA49 [7],
showing a $Q_{inv}$ correlation function and the distribution of
the event invariant radius.}
\end{center}
\end{figure}

\noindent
The first (preliminary) event-by-event analysis
of central $Pb+Pb$ collisions by NA49 [7] shows an approximately
symmetric distribution of the eventwise radius, the maximum well
corresponding to the ensemble average [8] of $R_{inv} = 7.5\:fm$. We
shall learn more about this signal from experiment STAR at RHIC
but note, for the time being, that the width of the distribution,
and even the slight indication of an asymmetry toward large
$R$ are roughly consistent with expectation from considering pair
counting statistics at fixed radius [9]; i.e. there seem to be
no prominent source volume fluctuations. At RHIC a more informative
analysis (in terms separately of longitudinal and transverse
source geometry) should prove feasible.

Related to HBT but much less developed theoretically is the geometrical
information derived from fireball nucleon coalescence into light
clusters. The deuteron or anti-deuteron coalescence factor [10]
$B_2=[d]/[p]^2$  is expected to fall down with increasing
source size and temperature (as well as with nucleon pair transverse
mass as shown by NA44 [11]), and the systematics has now reached
a perfect level by AGS E866 [12], E864 [13], SPS NA44 [11]
and NA52 [14]. However, completely striking was the vast extension
of coalescence physics by E864, both to heavier and metastable
nuclear species and to light hypernuclei [13,15]. Fig.~4 illustrates
their data for coalescence systematics up to $^7Be$, with yield
smaller by about 8 orders of magnitude than for the deuteron.
Normalizing away the overall exponential law there remains a
dependence on the cluster binding energy per nucleon (i.e. the
nucleon density). Will we next see exotic halo nuclei within
hot coalescence rather than cold fragmentation? E864 shows more
interest in the exciting physics of light hypernuclei
measuring the yield ratio of hyper-tritium to $^3He$ [15]. This
entire field is special to the AGS program.

\begin{figure}
\begin{center}
\epsfig{figure=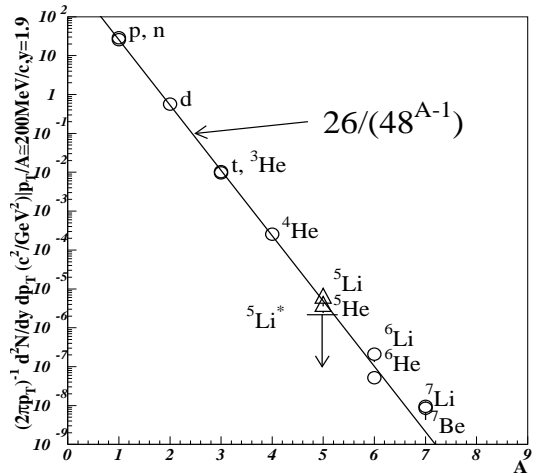,height=7cm, width=7cm}
\caption{Coalescence yield systematics by E864 [13] for $Au+Au$
at 11 $GeV/A$.}
\end{center}
\end{figure}

\section{Evolution with centrality and $\sqrt{s}$}

A decade of experimental search for the dependence of charmonia
($J/\psi$ and $\psi^*$) yields on system size and energy density
[16] has made us aware of threshold effects modifying a smooth
or even constant yield of some species relative to the number
of colliding baryon participants, or the total transverse energy
in the colliding system. Search for such changes with increasing
energy density was motivated by the QCD-Debey-screening model
of Matsui and Satz [17] which introduced discussion of several
characteristic steps in the energy density: if the transition
from hadronic matter to a quark gluon plasma was assumed at
$T=T_{crit}$ and corresponding energy density $\epsilon_{crit}$,
then the QCD screening mechanism dissolves first the $\psi^*$
and $\chi$ states, at $T \approx 1.1 \:T_c$,
then $J/\psi$ at $T \:\approx \:1.3\:T_c$
and, much later, the bottonium $Y$ at about 2.5 $T_c$. Thus
in this highly idealized, schematic model there is a sequence
of characteristic thresholds; the $J/\psi$, in particular,
would disappear at $\epsilon \approx 2.8 \:\epsilon_{crit}$ as $\epsilon$
is proportional to $T^4$ from lattice QCD [18]: i.e. \textbf{well above
the critical energy density}! At this conference the NA50 Collaboration
has shown a dramatic final drop in the $J/\psi$ yield occuring
at the most central $Pb+Pb$ collisions at
$\sqrt{s} = 17 \:GeV$. Let us for a moment accept that this
signals $J/\psi$
screening, and that this occurs at
$\epsilon \: \approx \:3 \: \epsilon_{crit}$
(in spite of the very intense ongoing discussion). Recall the
Bjorken estimate of the average (transverse) energy density in
the primordial reaction volume: for central $Pb+Pb$ it is
$\epsilon \: \approx \:3 \:GeV/fm^3$ from NA49/WA87 calorimetry [19].
This finally leads
to the estimate $\epsilon_{crit} \: \approx \: 1 \:GeV/fm^3$, in accord
with recent lattice QCD results [20]. This line of argument certainly
needs refinement at every step but may serve as a qualitative
guideline at least. We would then conclude that a threshold
at central $Pb+Pb$, and top SPS energy is characteristic for the
$J/\psi$ signal specifically, and that the conditions for
the hadron to parton phase change, i.e.
$\epsilon =\epsilon_{crit}$,
must be reached at more peripheral $Pb+Pb$ collisions, or in central
mass 200 collisions at lower $\sqrt{s}$. Insofar as hadronic observables
are sensitive
to the phase change they should exhibit a step, kink or such,
roughly speaking between $p+p,\: p+A$ or $^4He$ + $^4He$ (the former ISR
data [21]), and central $Pb+Pb$ at $\sqrt{s} = 17 \: GeV$.

\subsection{Hadron yields}

\begin{figure}
\begin{center}
\epsfig{figure=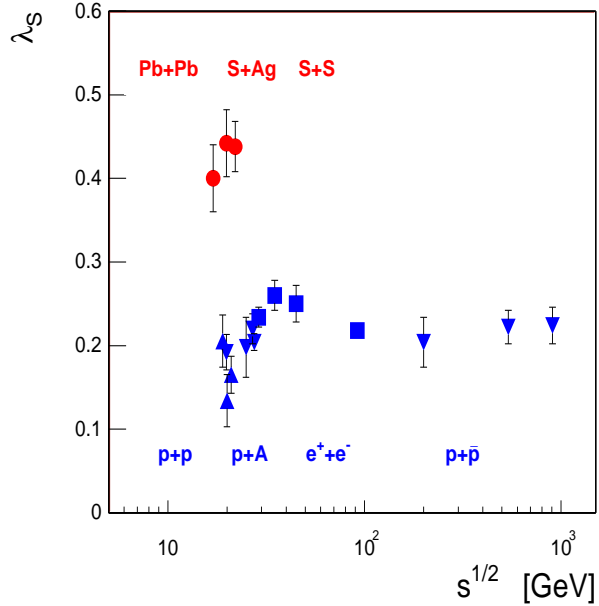,height=8.5cm, width=8.5cm}
\caption{Systematics of the Wroblewski strangeness suppression factor
$\lambda_s$ in elementary collisions and in central nucleus-nucleus
collisions at SPS energy [22].}
\end{center}
\end{figure}

Strangeness production data exhibit such an evolution. Fig.~5
shows the systematics of the so-called Wroblewski ratio
$\lambda\:=\:2<s+\overline{s}>/(<u+\overline{u}>+<d+\overline{d}>)$
 of
global strange to nonstrange quark content in 4$\pi$ obtained
by Becattini et al. [22] from an analysis of central $Pb+Pb$,
$S+Ag$ and $S+S$ collisions at the SPS, and from $p+p$, $p+A$, $e^+$  +
$e^-$ and $p + \overline{p}$. There occurs a jump of about two.
This enhancement
is most dramatically exhibited by the recent WA97 results concerning
hyperon production [23], see Fig.~6 (top) in $Pb+Pb$, with jumps
increasing in magnitude with
 $|s| \:=\: 1, 2, 3$ in the midrapidity yield relative to $p+Be$ (confirmed
by NA49 for 4$\pi$ cascade hyperon yields [24]). These enhancements
stay constant at $\langle N_{part}\rangle > 100$: the transition
must be below. The full range of impact parameters is, however,
covered in the same reaction for the relative $K/\pi$ yield
(i.e.($K^++ K^-) /(\pi^+ + \pi^- $))
by NA49 [25], and for the $\phi$ yield per participant by NA50
[26]: see Fig.~6 (bottom). Here we see a smooth increase from
$pp$ upwards to $N_{part} \approx 250$ where the ratios flatten out.
NA49 shows [27] that overall the $\phi$ to participant ratio in
full 4$\pi$ increases by a factor of 3.2 whereas
the $\langle \phi \rangle / \langle \pi \rangle$ ratio increases by
2.8, and the $\langle K \rangle /\langle \pi \rangle$ ratio by
 2.1.

\begin{figure}[t]
\begin{center}
\epsfig{figure=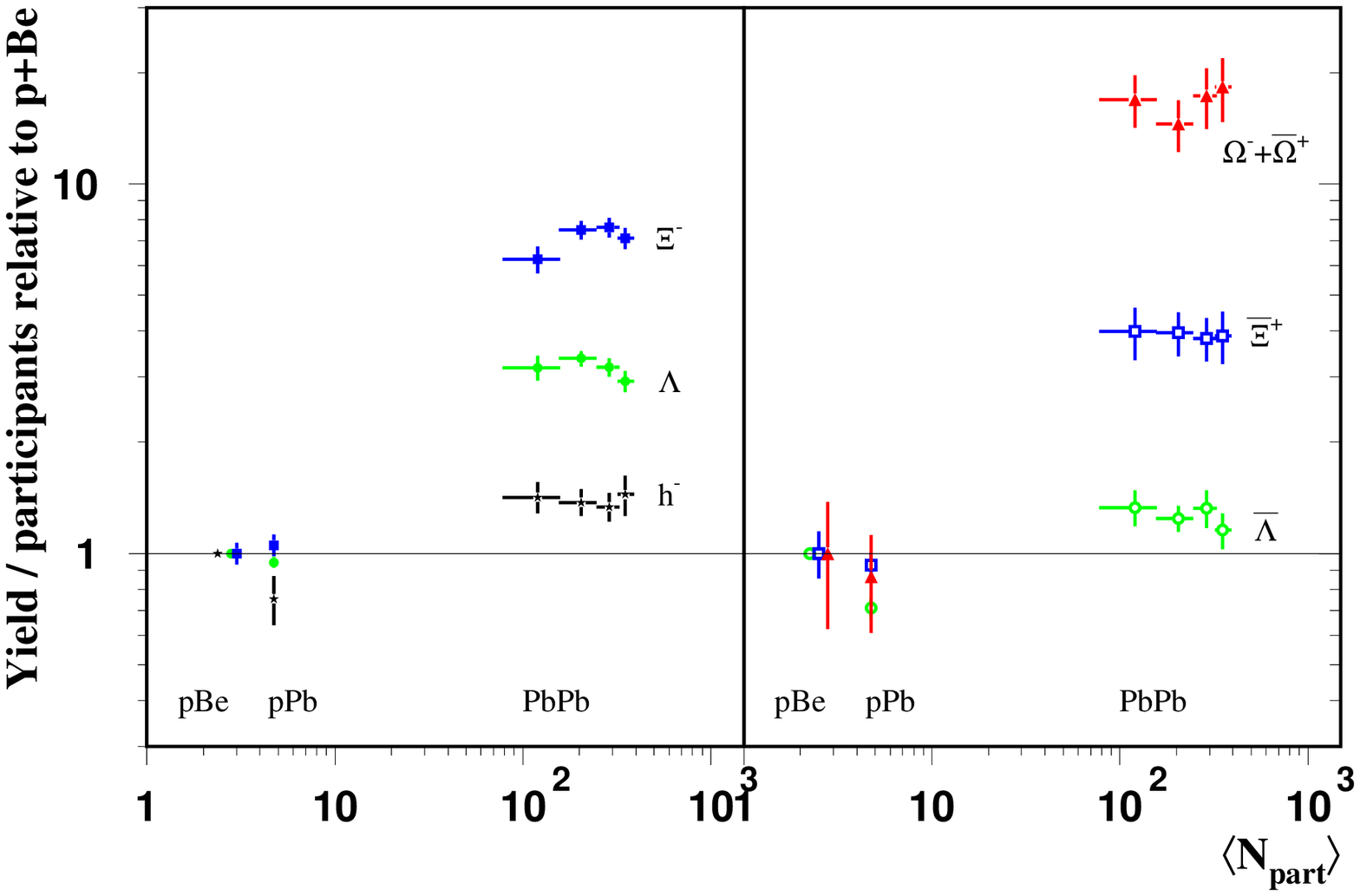,height=6.3cm, width=6.3cm}\\
\end{center}
\begin{minipage}[t]{5cm}
\epsfig{figure=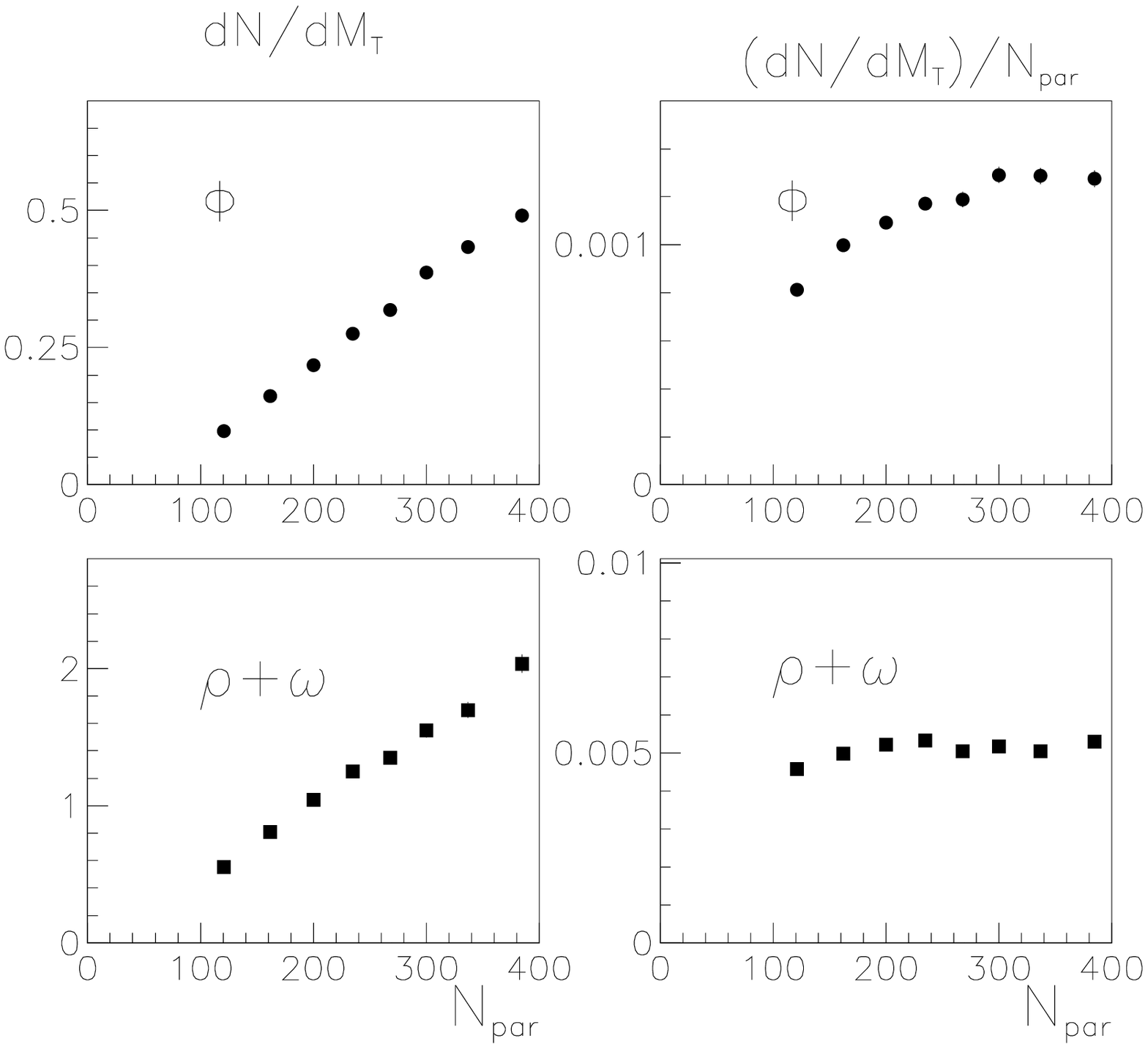,height=5.3cm, width=5.3cm}
\end{minipage}
\hspace{2cm}
\begin{minipage}[t]{5cm}
\epsfig{figure=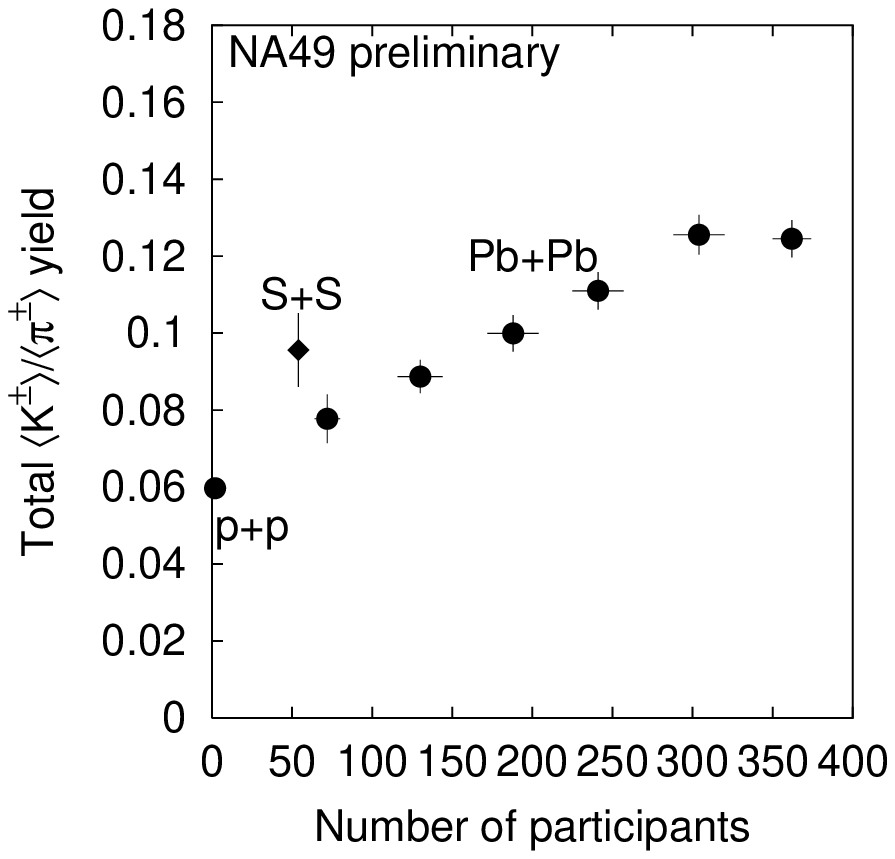,height=5.3cm, width=6cm}
\end{minipage}
\caption{Evolution of hadron production yields with centrality in $Pb+Pb$
collisions at 158 $Gev/A$. Hyperon data from WA97 [23], vector meson data
from NA50 [26] and the $K/\pi$ ratio from NA49 [25].}
\end{figure}

In contrast the nonstrange yields per participant exhibit
minor increases only: $\pi/N_{part}$ and $(\rho\:+\:\omega)/N_{part}$
by about 20\% [25,26]. NA49 sees a $\langle \overline{p} \rangle / N_{part}$
ratio that
is constant [28]. Returning to the $J/\psi$ yield within this
systematics Gazdzicki has shown [29] that the $\psi/\pi$
ratio is constant (about 10$^{-6}$) from $p+p$ up to minimum bias
$Pb+Pb$ ($\langle N_{part} \rangle \approx $ 100). NA50 might furnish the
continuation of this ratio toward the most central collisions
(down by a factor of about 2 to 3 ?).

Overall, the systems of hadronic yield ratios undergoes a dramatic
change in going from $p+p$, $p+A$ to central $Pb+Pb$ at
$\sqrt{s}\: =\: 17 \:GeV$; the changeover occurs at semi-central $Pb+Pb$ and
below central $S$ + heavy nucleus (the data are not as comprehensive).
At the AGS the trend is similar but less dramatical [30] perhaps
due to the lack of hyperon yield systematics. We will return
to the hadronic composition of the final state in section 6,
showing that it holds the key to determine the parameters of
the parton to hadron phase transformation.

\subsection{Bose-Einstein correlations}

Data of high precision are now available for the evolution of
HBT ''geometrical'' source parameters with system size (centrality),
$\sqrt{s}$ and pair transverse invariant mass, both from the AGS and the
SPS experiments. Concerning the evolution with system size the
AGS E802/866 Collaboration showed [31] that the para\-meter
$R_{side}$
(which lends itself most directly to characterize the transverse
source dimension [4]) is very strictly proportional to the cube
root of the participant nucleon number from peripheral $Si+Al$
to central $Au+Au$ collisions. A result augmented by NA49 [32]
which considers $R_{side}$ as a function of the cube root of the pion
multiplicity density near midrapidity for the sequence $p+p$, $p+Pb$,
$S+S$ and $Pb+Pb$ at various impact parameters. These data are shown
in Fig.~7.

\begin{figure}
\begin{center}
\epsfig{figure=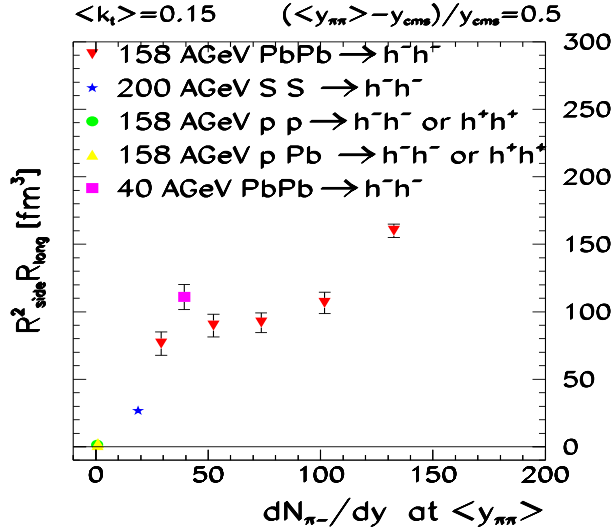,height=7cm,width=8cm}
\caption{Systematics of the HBT eigenvolume $R^2_{side}R_{long}$ with
negative pion rapidity density near midrapidity for $p+p$, $p+Pb$,
$S+S$ and $Pb+Pb$ by NA49 [32].}
\end{center}
\end{figure}

They include
the first result from the SPS trial run at 40 $GeV/A$ for central
$Pb+Pb$, the data point sitting somewhat high within the surrounding
$R_{side}$ values for semiperipheral $Pb+Pb$ at 158 $GeV/A$. A feature
reminiscent of the similar excursion in the $K/\pi$ ratio
(Fig.~6 bottom) exhibited by the central $S+S$ data point falling
above the systematics of participant number dependence in $Pb+Pb$.
A possible hint that neither the participant number scale nor
the midrapidity density of produced particles (pions) properly
reflects the difference of a central collision, at lower $A$ or
$\sqrt{s}$, and the seemingly concurrent, fairly peripheral $Pb+Pb$
collisions. One might suspect the higher net baryon density near
midrapidity, or the geometrically more compact primordial interaction
volume of central collision to cause this second order effect which
deserves further study.

AGS E866 also reports [31] comprehensive data concerning
the ''duration of emission'' parameter. In terms of the
Bertsch-Pratt
variables it is given by the square root of $R_{out}^2 - R_{side}^2$ and
denotes the temporal width of the source luminosity with regard
to decoupling pion pairs. It approaches 4 $fm/c$ in central $Au+Au$,
in close agreement with the former NA49 analysis [33] of central
$Pb+Pb$ at the SPS. This parameter highlights the wider informational
content of HBT analysis. Beyond the mere spatial geometry of
the source, traditionally expected from such data, it elucidates
the time scale of source decomposition upon expansion which thus
turns out to be very short owing to an ''explosive'' hadronic
expansion dynamics [34]. Actually it is equally short at
$\sqrt{s}=4.3$, 8.8 and 17 $GeV$ [32].
We shall return to this information in section 5, for the discussion
of global hadronic freezeout.

Whereas we see a smooth increase of $R_{side}$ and $R_{out}$ with centrality
the NA49 minimum bias data [32] indicate a flattening out of
the longitudinal radius parameter which is related to the overall
reaction time. Like the duration of emission parameter it shows
no variation at 40 $GeV/A$ [32] but is much smaller at AGS $Au+Au$ at
11 $GeV/A$ [31], a somewhat counterintuitive finding. In
any event the two reaction time related parameters indicate no
maximum anywhere, including the lower AGS energies 2 to 8 $GeV/A$
that were analyzed by E895 [35]. They thus give no reason to
suspect that the ''softest point'' of the EOS (i.e. the ''position''
of the hadron-parton phase transition) has been hit which might
reflect in an extended time scale [35,36] due to existence of a
long-lived mixed phase. This brings me to the topic of the next section.

\section{Evolutions and onset searches in directed flow observables}

The occurence of directed flow phenomena can be linked to the
time integral over accelerating forces, carried out over the
entire expansions dynamics until freezeout. The forces relate
to unisotropic local pressure fields, the pressure in turn results from density
and temperature via the equation of state (EOS). Rischke [37]
and Shuryak [38,39] have shown repeatedly how various models
of a phase transition between a hadronic resonance gas and a
schematic quark gluon phase exert their influence via the EOS
on directed ($\nu_1$) and elliptic ($\nu_2$) flow of various
hadrons. These schematic models focus on the transition ''point''
 resulting in a mixed phase which causes the pressure
to fade: the ''softest point'' of the dynamics which causes a
discontinuity in the excitation function of the observable at
the ''onset'' of the new phase. It has often been remarked that
the terminology of points and onsets might be misleading as each
colliding system features large gradients of energy density
 in its interior sections: there should be no sudden or
pointlike phenomena if a realistic hydrodynamical model is consulted.
In fact Schlei [40] reported such calculations with the HYLANDER
model which did not confirm sudden effects upon (gradually) crossing
the EOS boundary.

Two sets of data emerged as possible candidates for a smooth onset
behaviour. The beautiful systematic AGS data, combined from E877
[41] and EOS [30,43] $Au+Au$ provide for an excitation
function of directed and elliptic flow, the latter changing sign
at about 4 $GeV/A$. Danielewicz [44] showed that $\nu_2$ may
be best reproduced by a transition from a hard ($K=380\: MeV$)
to a soft ($K=210 \:MeV$) EOS, occuring between about 2 and 5 $GeV/A$.
A si\-milar smooth effect was pointed out in the NA49 elliptic
pion flow data [45] for minimum bias $Pb+Pb$ by Sorge [46] employing
RQMD with the microscopic adaptation of a phase transition which
leads to a kink in $\nu_2$  at impact para\-meters from
about 6 to 9 $fm$, not unlike the data. But we are now faced with onset
candidates  at $E=4 \:GeV/A$ and 158 $GeV/A$, in semi-central
mass 200 collisions.

Flow of strange particles offers a different window on the EOS
of dense matter. For kaon directed in plane flow we have
data at 1-2 $GeV/A$ from SIS [47] and were presented new AGS data
for $Au+Au$ shown in Fig.~8: Neutral kaons
$K^0_s$  exhibit a dramatic $\langle {p_x} \rangle$ signal in the 6 GeV/A E895 data
[30] (top) perpendicular to the proton flow (and in striking
disagreement with the RQMD (2.3) $K^0$ prediction) thus pointing
to a strongly repulsive in-medium nucleon neutral kaon potential
[48] which is baffling because
$K^0_s$ is $K^0$ and  $\overline{K^0}$ to equal parts. The E877
data [42,49] are for the
top AGS energy 11.5 $GeV/A$ and show (bottom Fig.~8) that both
$K^+$ and $K^-$ show very little $\langle {p_x} \rangle$ throughout (like at SIS
energy). As such they also push away from the protons and lambdas
but $K^0_s$ is really anti-correlated to the latter.

\begin{figure}
\begin{center}
\epsfig{file=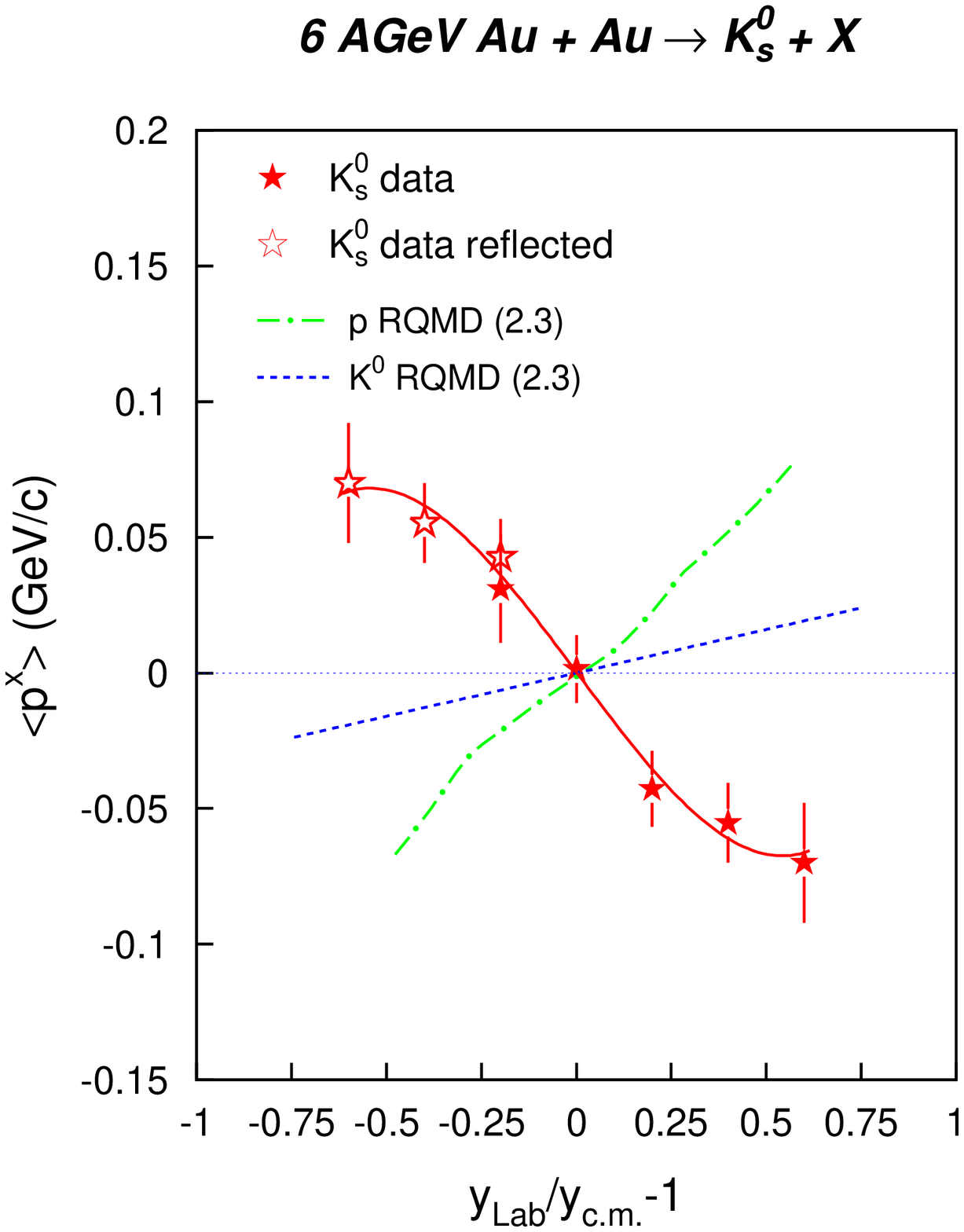,height=6cm,width=6cm}
\epsfig{file=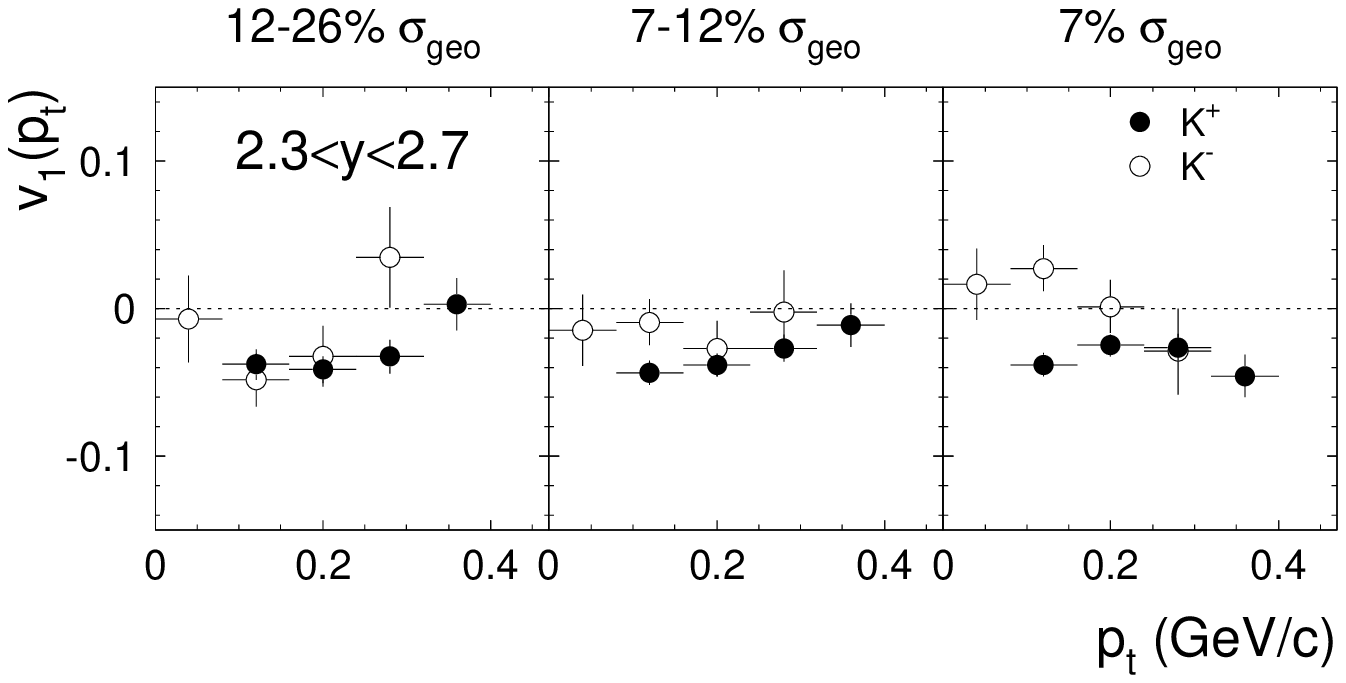,height=6cm,width=11cm}
\caption{In plane directed flow of $K_s^0$ as function of $cms$ rapidity
in semiperipheral $Au+Au$ at 6 $GeV/A$ by E895 [30] (top) and
$p_t$ dependence of $K^+,\:K^-$ directed flow at 11 $GeV/A$ by E877 [42].}
\end{center}
\end{figure}

\section{Global freezeout and radial velocity fields}

In considering the final phase of hadronic fireball expansion we first focus
our attention at the global freezeout: when hadrons decouple
from strong interaction (with exception of a few resonances left
over which decay even later, during free-streaming). It is
known [33,50] that collective velocity fields, both in transverse
and (even stronger) in longitudinal direction, dominate the hadronic
transverse mass spectra and the pair correlations (in their dependence
on $m_T$). This so-called hadronic radial flow seems at least in
part to result from the near isentropic nature of hadronic expansion
[6,51], and may have contributions from the pressure during the
preceding partonic phase if it exists.

All of this has been discussed already at QM97 but we have new insight
into the inverse $m_T$ slope systematics previously established
for SPS central $Pb+Pb$ collisions by NA44, NA49 and WA97 [50,52,53].
We illustrate this in Fig.~9.

\begin{figure}
\begin{center}
\epsfig{figure=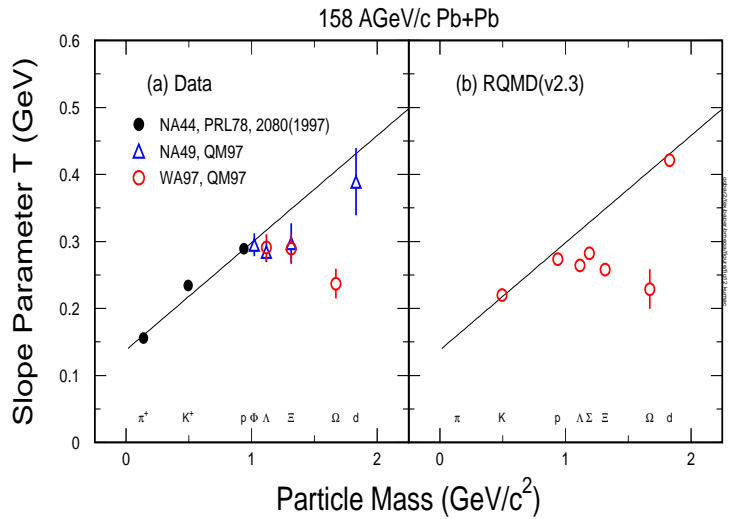,height=8cm, width=10cm}
\caption{Systematics of hadron inverse slope parameters from pion to
deuterium for central $Pb+Pb$, and RQMD results by Nu Xu [54] showing
hyperon ''excursion''.}
\end{center}
\end{figure}

Nu Xu showed [54] that RQMD (2.3)
accounts well for the previously noted excursion of the hyperon
$m_T$-spectral slope parameters, notably for the cascade and omega
particles, from the linear increase of $T$ with hadron mass that
would result if all hadrons froze out simultaneously from a common
radial velocity field. But Nu Xu showed that e.g. the departure
of the $\Omega$ slope results in RQMD from an early decoupling
owing to the small total cross section. After emerging from hadronization
the $\Omega$ scatters only about once or twice whereas protons
undergo about 6 successive, mostly elastic collisions [55]. This
offers the tantalizing prospect to determine with low cross section
hadrons the transverse flow existing directly near the stage
of primordial hadrosynthesis, at which the true temperature is
about 170-180 $MeV$ as I shall describe in the next section [22,56,57].
The observed $\Omega$ slope $T \:=\: 230\: MeV$ would
then give an estimate
of ''primordial flow'', from the partonic phase, amounting to
$\beta_{\bot}\:=\:0.27$ by
using the Heinz formula [58] $T=T_{true} \sqrt{(1 + \beta)/(1 - \beta)}.$
Next to the $\Omega$
the $\phi$ would be most interesting in this regard but NA49
and NA50 report a discrepancy here: $T = 295$ and 225 $MeV$, respectively
[26,27].

To determine the freezeout conditions ($T_{true}, \:\beta_{\bot}$)
for the highly abundant hadrons $\pi, \:K, \overline{p}$ Sollfrank
and Heinz [58] suggested long ago to resolve the
$T, \:\beta_{\bot}$ ambiguity
of each individual fit to the transverse mass spectrum by superimpo\-sing
the three ''corridors'' of $T, \:\beta_{\bot}$ and looking for
their overlap in the $T, \:\beta_{\bot}$ plane which they found
at $T_{true} = 125 \: MeV$ and $\beta_{\bot} = 0.6$ for SPS
sulphur
collision data. Note that this procedure does \textbf{not} pick out
the minimum $\chi^2$ fit values of $T,\:\beta_{\perp}$
for individual hadrons which might be \textbf{different} as Peitzmann
[1] demonstrated for the neutral pion, wide $m_T$ range WA98 data
(shown in Fig.~1), also pointing out all the difficulties arising
in the pion spectra from resonance feeding, relativistic flow
and Croonin effect contributions. It seems advisable to restrict
to kaon, proton, antiproton and deuteron spectra in order to pin
down the global freezeout conditions.

NA49 has suggested [33] an alternative method combining spectral
data with HBT analysis because the hydrodynamically inspired source
model [6,51] predicts the transverse HBT radius parameter to fall
off with $m_T$ of the pair as
\begin{equation}
R_T(m_T)=R_G(1+\frac{m_T\beta^2_{\bot}}{T})^{-1/2}
\end{equation}

\noindent
at midrapidity; here $R_G$ is the ''true'' transverse radius of the
source, and $T$ is the ''true '' freezeout temperature. The fit of
the $m_T$ dependence of $R_T$ then contains another continuous
ambiguity, this time $R_G$ vs. $\beta^2/T$ or
$\beta^2$ vs. $T$ if $R_G$ gets fixed. Fig.~10
illustrates this procedure [33] for central $Pb+Pb$ collisions.
The information from spectra (which implicitly results from
total transverse energy conservation [59]) and HBT (transverse coherence
length conservation [60]) comes at right angles. NA49 concluded
$T=120\:MeV$, $\beta_{\bot}$ = 0.55 but in the light of the above discussion
one might now ignore the negative hadron (i.e. mostly $\pi$)
information deducing the bulk hadron \textbf{freezeout conditions}
$T =110\:\pm \:10 \:MeV$, $\beta_{\bot}=0.60\:\pm\:0.08$
for central $Pb+Pb$ near midrapidity in agreement with
Wiedemanns [4] analysis.

\begin{figure}
\begin{center}
\epsfig{file=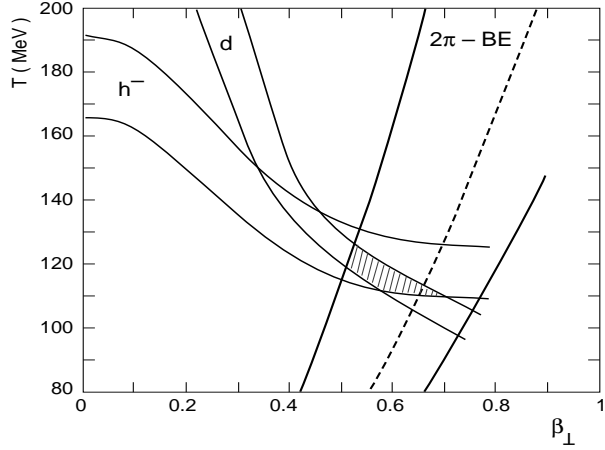,height=6cm,width=8cm}
\caption{Global freezeout parameters $T$ and $\beta_{\bot}$ for
central $Pb+Pb$ combining $m_T$ spectral information from negative
hadrons and deuterons with pion correlation analysis by NA49 [33].}
\end{center}
\end{figure}

The crucial correlation data that enter this quantification of
a very strong ''flow'' phenomenon can be expanded as NA44 showed
[61]. Fig.~11 reproduces their compounded data for the pair $m_T$
dependence of pion, kaon and proton two particle Bose (Fermi)
correlation and from deuteron coalescence, all for $Pb+Pb$ at
$\sqrt{s}$ = 17 $GeV$. First data on   the $m_T$ dependence of HBT radii
in $p+p$ and $p+Pb$ at similar energy were shown by NA49 [32]: some
effect is visible (not surprisingly) in the longitudinal radius
but the data are inconclusive about the transverse radii at the
present level of statistics.
\begin{figure}
\begin{center}
\epsfig{file=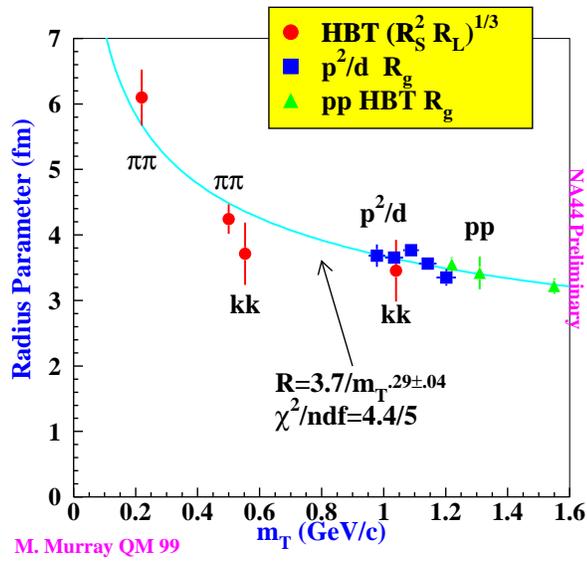,height=8cm,width=8cm}
\caption{Systematics of $m_T$ dependence for correlation radii from
$\pi\pi,\:KK,\:pp$ and deuteron coalescence in central $Pb+Pb$
collisions by NA44 [61].}
\end{center}
\end{figure}

Finally: when does this final freezeout occur? There ought to
be a wide spread in decoupling times as the surface of the primordial
interaction volume starts emitting hadrons into free streaming
right after formation time. From pion pair HBT analysis in the
framework of the expanding source model [6,51] one extracts two
relevant parameters: the average decoupling time $\tau$ and
the Gaussian duration of emission width $\Delta \tau$ (see
previous section). For central $Pb+Pb$ NA49 deduced [33,56] the
values
$\tau$ = 8 $fm/c$ and $\Delta \tau$ = 4 $fm/c$
at midrapidity.
Decoupling would thus cease at about 20 $fm/c$. Sorge [46] has
challenged the validity of the thus derived time scale and, in
fact, both the RQMD [62] and VNI models [63] put the end of the decoupling
phase to about 30 $fm/c$, in this collision. This needs further
discussion.

\section{Primordial hadrosynthesis and \-the parton-ha\-dron phase transformation}

If we, tentatively, accept the hypothesis [64] that the steep drop
of the $J/\psi$ yield in the most central $Pb+Pb$
collisions reported at this Conference by NA50 [65] signals a
partonic phase which Debeye-screens this tightly bound hadron we
might conclude that the average energy density amounts to about 3
times the critical density $\epsilon_c$ of the QCD deconfinement
transition. In section 3 we have hence argued that
$\epsilon_c \approx  1 GeV/fm^3$ invoking the Bjorken estimate
$\epsilon = 3 \:GeV/fm^3$ obtained for central $Pb+Pb$ collisions
[19]. One then expects to receive
signals specific to the exit from this phase, i.e. from the bulk
parton to hadron phase transformation bound to occur once the
primordial high energy density state expands and cools toward
$\epsilon_c$ and $T_c$. Determination of these critical parameters
is the principal goal of our research (perhaps adding the need to
specify the order of this phase transformation).

In this section I shall argue that the set of bulk hadronic production
rates and ratios is a direct consequence of this parton to hadron
phase transformation, occuring in central $Pb+Pb$ collisions at top
SPS energy, and perhaps already in central $S$ + mass 200 collisions.
In spite of the evidence discussed in the last sections for an
extensive dynamical evolution and collective expansion of the
fireball I argue that its chemical composition is hardly changed
by \textbf{inelastic} hadronic final state interactions [55]. The observed
hadron yields thus reflect the primordial conditions at the parton
to hadron phase transition, i.e. the critical values  $T_c$ and $\epsilon_c$
and, moreover, the flavour composition at the end of the partonic
phase. This view [56] was suggested by hadrochemical model stu\-dies
[22,58,66] of hadron production ratios, as confronted by models
of parton hadronization [63,67]. It was extensively  discussed by Heinz
at this Conference [57]. It is based on a coherent analysis of three
seemingly separate investigations of hadron production
data.

Firstly, Ellis and Geiger [68] developed in 1996 a partonic
transport model for the analysis of LEP data concerning
W and Z decay to hadrons at
$\sqrt{s} \approx 90 \:GeV$. The model (known now as VNI) consists of
an initialization step creating an initial quark-gluon population
that is evolved in a QCD partonic transport mode which, in turn,
ends in a coalescence phase creating colour singlet pre-hadronic
resonances that decay into the observed hadrons. Good overall
agreement was obtained with the data comprising all hadronic yields
up to hyperons. Most importantly the authors noted a remarkable
insensitivity of the yield distribution to the detailed assumptions
made in the model concerning the non-perturbative mechanisms
of hadronization. They concluded that the statistical phase space
weights of the various hadronic species overwhelm all other microscopic
influences, and that the final multi-hadronic population distribution
thus represents the \textbf{maximum entropy} state of highest probability.

Secondly, it is therefore not surprising that thermal hadro-chemistry
mo\-dels of Hage\-dorn-type successfully describe the same LEP data. Becattini
[69] employed a canonical partition function showing that the
multihadronic state represented a ''hadrochemical family'' characterized
by only two essential para\-meters, a temperature $T = 165\: MeV$ and
a strangeness under-saturation factor $\gamma_s$ = 0.65. As there
is no inelastic hadronic rescattering whatsoever in such
Z, W decays it was obvious that such an ordered state of hadrons
could not result from any equilibrating interaction at the hadronic
side. The statistical order is created \textbf{from above,} i.e. from
higher energy density prevailing at the onset of hadronization,
at the critical QCD energy density $\epsilon_c$. Arriving at the
hadronic side of the phase transformation this state is partially
out of proper hadronic chemical equilibrium as indicated by
$\gamma_s \:<\:1$.
The preceding partonic phase enters its specific strange
quark/antiquark population (which may or may not represent a
flavour equilibrium at the partonic level) into the hadronization
process which creates essentially no new strangeness but plenty
of light quarks. The emerging hadronic population thus appears
strangeness-undersaturated
at the transition temperature, and it can not readjust the strangeness
population owing to its own proper ''explosive'' expansion [55,56]. It
thus preserves the signatures of the phase transition, which
freeze out at the instant of hadronization.

Thirdly, it turns out that the multihadronic population order
observed in central $Pb+Pb$ collisions can be understood similarly
as a fingerprint of the parton to hadron phase transformation
[56,57,66,67,69,71]. However, we must expect significant differences in
detail proceeding from the outcome of LEP $e^+ + e^-$ data analysis to
central $Pb+Pb$ collisions. Recall that the hadronic population
ratios were shown in section 3 to exhibit a dramatic evolution
in proceeding from $p+p$ via minimum bias $Pb+Pb$, to finally settle
into a new pattern becoming stationary from semi-central collisions
onward (Fig.~6) to central $Pb+Pb$ at the top SPS energy. There
seems to occur a transition from hadron population patterns characteristic
of elementary $e^+ + e^-$,  $p+p$, $p + \overline{p}$ collisions at
$\sqrt{s} \geq  20 \:GeV$ to central $Pb+Pb$ at the same energy.
The hadrochemical
model analysis reflects this transition in its proper parameters
$T$, $\gamma_s$ and baryochemical potential $\mu_B$ but again
leads to a perfect fit of the hadronic production ratios.

\begin{figure}
\begin{center}
\epsfig{file=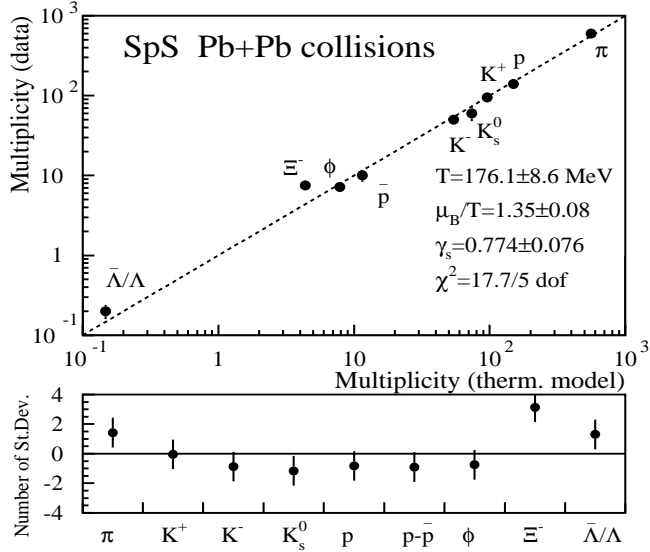,height=8cm,width=9cm}
\caption{Grand canonical thermal model analysis of hadron yields
and yield ratios in central $Pb+Pb$ collisions by Becattini [70].
Preliminary data from NA49 [25].}
\end{center}
\end{figure}

Figs.~12 and 13 illustrate the recent analysis of $Pb+Pb$ central
collision data by Becattini [70] and by Braun-Munzinger et
al. [66]. The former analysis refers to NA49 data only whereas
the latter includes an interpolation of all available data from
NA44, NA49 and WA97. Both models employ the grand canonical version
of the thermodynamical equilibrium model but differ in detail (concerning
employment of $\gamma_s$ and hadron eigenvolume corrections).
However the outcome agrees rather closely:
Becattini derives $T = 176\:\pm$ 9 $MeV$,
$\mu_B = 237 \:\pm 20 \:MeV$, $\gamma_s = 0.77 \pm $ 0.08 and
Braun-Munzinger et al. get $T = 170 \:\pm$ 10 $MeV$,
$\mu_B = 270 \:MeV$ and $\gamma_s$ = 1 (the latter a fixed parameter in
this analysis). Thus, in comparison to the $e^+ + e^-$ hadronization
results, the temperature of primordial hadronization may be slightly
higher, the baryochemical potential (near zero in elementary
 collisions) acquires prominence by the presence of an excess
of light quarks over light antiquarks, due to the stopped-down initial high
valence quark density in the participating nuclear matter. Most
prominently the strangeness undersaturation, characteristic
of elementary collisions (Fig.~5) fades away. This is not resulting
from any inelastic equilibration at the hadronic side of the
phase transformation which does not take place owing to the ''explosive''
expansion cooling of the interaction volume as Heinz has shown
[55,56,57]. The hadronic expansion mode is governed by elastic interactions
which produce the near isentropic ''flow'' mode generating the
collective velocity fields that I described in section 5. The
hadronic composition thus freezes out immediately at the instant
of hadronization, the chemical composition being out of thermal
equilibrium ever after. The apparent fading of the strangeness
suppression factor $\gamma_s$ (which leads to the various levels
of strange hadron yield enhancement, in comparison to elementary
collisions) must, therefore, stem from the \textbf{partonic} phase
flavour balance ( as was demonstrated by Zimanyi [71] at this
conference), entering a higher primordial strangeness saturation
level into the hadronization transition. In fact Sollfrank
et al. [72] have shown that the observed $s + \overline{s}$
density is compatible
with hadronization from a flavour equilibrated quark-gluon plasma
at $T \approx 180\:MeV$.
\begin{figure}
\begin{center}
\epsfig{file=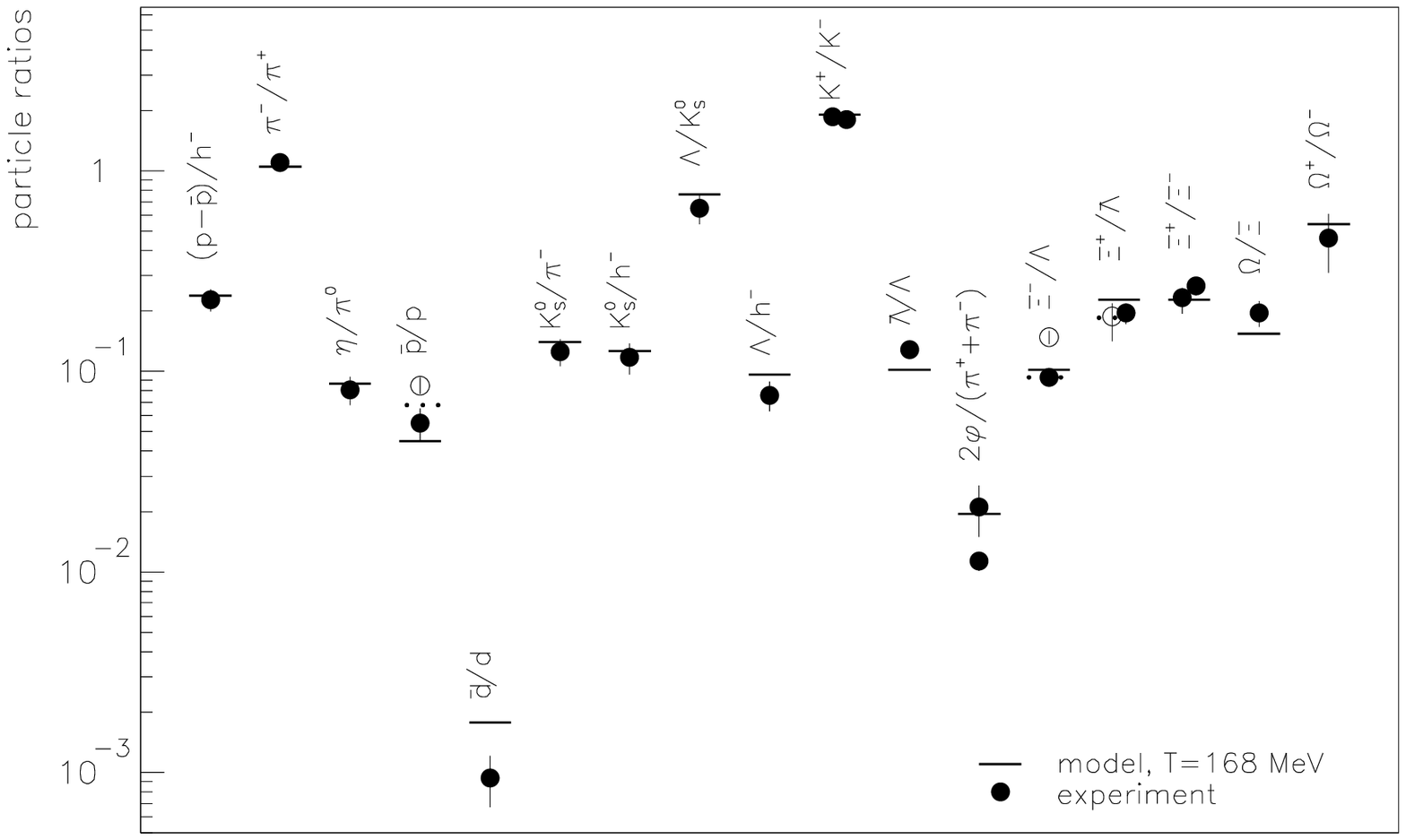,height=8.5cm,width=8.5cm}
\caption{Thermal model analysis by Braun-Munzinger et al. [66]
for central $Pb+Pb$ hadron yields compiled from NA44, 49 and WA97.}
\end{center}
\end{figure}

We thus seem to narrow in onto the most cherished topic of the
field: pinning down the parameters of the parton to hadron phase
transformation. This was the highlight of QM99 as far as hadronic
signals are concerned. As for a summary of the considerations
that lead to this conclusion I chose to give U. Heinz [57]
the last word: ''The analysis of soft hadron production data at
the SPS indicates that hadron formation proceeds by statistical
hadronization from a prehadronic state of deconfined quarks.
This leads to pre-established apparent chemical equilibrium among
the formed hadrons at the confinement temperature $T_c$; it is not
caused by kinetic equilibration through hadronic rescattering.
After hadronization the hadron abundances freeze out more or less
immediately. The chemical freeze out temperature thus coincides
with the critical QCD temperature,
$T_{chem} \approx  T_c  \approx  170-180 \:MeV$
(at $\mu_B \approx  250\: MeV$), corresponding to a critical
energy density $\epsilon_c \approx 1 \:GeV/fm^3$ as predicted
by lattice QCD''.


\begin{thebibliography}{99}
\bibitem{1} Th. Peitzmann, these Proceedings;\\
Ch. Blume, PhD Thesis, Univ. M\"unster (1998).
\bibitem{2} B. Lenkeit, these Proceedings.
\bibitem{3} M. Gyulassy, these Proceedings.
\bibitem{4} U. A. Wiedemann, these Proceedings.
\bibitem{5} M. Lisa, these Proceedings.
\bibitem{6} S. V. Akkelin and Yu. M. Sinyukov, Phys. Lett. B356
(1995) 525; \\
U. A. Wiedemann, P. Scotto and U. Heinz, Phys. Rev.
C53 (1996) 918; \\
T. Cs\"org\"o and B. L\"orstad, Phys. Rev. C54
(1996) 1390.
\bibitem{7} G. Roland, private communication.
\bibitem{8} H. Appelsh\"auser, PhD Thesis, GSI,DISS 97-02
\bibitem{9} D. Ferenc, CERN ALICE note 93-14; \\
R. Stock, Proc. NATO Divonne Workshop 1994, Editors J. Letessier, H. Gutbrod and
J. Rafelski.
\bibitem{10} A. Mekjan, Phys. Rev. Lett. 38 (1977) 640.
\bibitem{11} I. Bearden, these Proceedings.
\bibitem{12} E866 Collaboration, private communication.
\bibitem{13} G. van Buren, these Proceedings.
\bibitem{14} G. Ambrosini et al., NA52 Coll., Nucl. Phys. A638
(1998) 411c.
\bibitem{15} E. Finch, these Proceedings.
\bibitem{16} L. Kluberg, these Proceedings.
\bibitem{17} T. Matsui and H. Satz, Phys. Lett. B178 (1986) 416.
\bibitem{18} E. Laerman, Nucl. Phys. A610 (1996) 1c.
\bibitem{19} T. Alber et al., NA49 Coll., Phys. Rev. Lett. 75 (1975)
3814;\\
M. Aggarwal  et al., WA98 Coll., Nucl. Phys. A610 (1996) 200c.
\bibitem{20} F. Karsch, Nucl. Phys. A590 (1995) 367c.
\bibitem{21} H. G. Fischer, Z. Phys. C38 (1988) 105.
\bibitem{22} F. Becattini, M. Gazdzicki and J. Sollfrank, Nucl.
Phys. A638 (1998) 403c.
\bibitem{23} E. Andersen et al., WA97 Coll., Phys. Lett. B433
(1998) 209;\\
F. Antinori, these Proceedings.
\bibitem{24} H. Appelsh\"auser et al., NA49 Coll., Phys. Lett.
B444 (1998) 523.
\bibitem{25} F. Sikler, these Proceedings.
\bibitem{26} N. Willis, these Proceedings.
\bibitem{27} C. H\"ohne, these Proceedings.
\bibitem{28} G. I. Veres, these Proceedings.
\bibitem{29} M. Gazdzicki and M. I. Gorenstein, hep-ph/9905515 to
be published in Acta Phys. Pol.
\bibitem{30} G. Rai, these Proceedings.
\bibitem{31} R. Solz, these Proceedings.
\bibitem{32} R. Ganz, these Proceedings.
\bibitem{33} H. Appelsh\"auser et al., NA49 Coll., Eur. Phys. J.
C2 (1998) 661.
\bibitem{34} J. Letessier, J. Rafelski and A. Tounsi, Nucl Phys.
A590 (1995) 613c.
\bibitem{35} S. Pratt, Phys. Rev. C33 (1986) 1314.
\bibitem{36} G. F. Bertsch, Nucl. Phys. A498 (1989) 173c.
\bibitem{37} D. H. Rischke, Nucl. Phys. A610 (1996) 88c.
\bibitem{38} E. V. Shuryak, Nucl. Phys. A638 (1999) 207c.
\bibitem{39} E. V. Shuryak, these Proceedings.
\bibitem{40} B. Schlei, these Proceedings.
\bibitem{41} J. Barette et al., E877 Coll., Nucl. Phys. A638
(1998) 69c.
\bibitem{42} K. Filimonov, these Proceedings.
\bibitem{43} H. Liu et al., E895 Coll., Nucl. Phys. 638 (1998)
451c.
\bibitem{44} P. Danielewicz, these Proceedings.
\bibitem{45} A. M. Poskanzer, these Proceedings.
\bibitem{46} H. Sorge, these Proceedings.
\bibitem{47} Y. Shin et al., KaoS Coll., Phys. Rev. Lett. 81
(1998) 1576.
\bibitem{48} G. Q. Li et al., Phys. Lett. B329 (1994) 149.
\bibitem{49} J. Barette, these Proceedings.
\bibitem{50} I. G. Bearden et al., NA44 Coll. Nucl. Phys. A638
(1998) 103c.
\bibitem{51} J. D. Bjorken, Phys. Rev. D27 (1983) 140;
U. Heinz, Nucl. Phys. A610 (1996) 264c.
\bibitem{52} H. Appelsh\"auser et al., NA49 Coll., Phys. Rev. Lett.
82 (1999) 2471.
\bibitem{53} D. Elia, these Proceedings;\\
R. Lietava et al., WA97 Coll., J. Phys. G25 (1999) 181.
\bibitem{54} Nu Xu, these Proceedings.
\bibitem{55} A. Dumitru et al., nucl-th/9901046;\\
S. Bass et al., nucl-th/9902062.
\bibitem{56} R. Stock, Phys. Lett. B456 (1999) 277.
\bibitem{57} U. Heinz, these Proceedings, nucl-th/9907060.
\bibitem{58} J. Sollfrank and U. Heinz, in Quark Gluon Plasma 2,\\
R. Hwa (Ed.), World Scientific.
\bibitem{59} R. Stock, Nucl. Phys. A630 (1998) 535c.
\bibitem{60} T. Cs\"org\"o and B. L\"orstadt, Nucl. Phys. A590
(1995) 465c.
\bibitem{61} M. Murray, these Proceedings.
\bibitem{62} H. Sorge, Phys. Lett. B373 (1996) 16.
\bibitem{63} K. Geiger, Nucl. Phys. A638 (1998) 551c; \\
K. Geiger and D. K. Srivastava, Phys. Rev. C56 (1997) 2718.
\bibitem{64} H. Satz, these Proceedings.
\bibitem{65} C. Cicalo, these Proceedings.
\bibitem{66} P. Braun-Munzinger, I. Heppe, J. Stachel,
nucl-th/9903010, to appear in Phys. Lett. B; \\
J. Stachel, these Proceedings.
\bibitem{67} W. Bartz, B. L. Friman, J. Knoll and H. Schultz,
Nucl. Phys. A519 (1990) 831.
\bibitem{68} J. Ellis and K. Geiger, Phys. Rev. D54 (1996) 1967.
\bibitem{69} F. Becattini, Z. Phys. C69 (1996) 485, and
hep-ph/9701275.
\bibitem{70} F. Becattini, private communication.
\bibitem{71} J. Zimanyi, these Proceedings; \\
J. Zimanyi, T. S. Biro and P. Levai, hep-ph/9904501.
\bibitem{72} J. Sollfrank, F. Becattini, K. Redlich and H. Satz,
Nucl. Phys. A638 (1998) 399c.

\end{thebibliography}
\end{document}